

\input harvmac


\overfullrule=0pt


\def\B{{\scriptscriptstyle B}}

\def\D{{\scriptscriptstyle D}}

\def\F{{\scriptscriptstyle F}}

\def\I{{\scriptscriptstyle I}}

\def\L{{\scriptscriptstyle L}}

\def\N{{\scriptscriptstyle N}}

\def\P{{\scriptscriptstyle P}}

\def\R{{\scriptscriptstyle R}}

\def\W{{\scriptscriptstyle W}}

\def\Z{{\scriptscriptstyle Z}}


\def\CA{{\cal A}}

\def\CD{{\cal D}}

\def\CL{{\cal L}}
\def\CM{{\cal M}}


\def\e{\epsilon}
\def\g{\gamma}

\def\o{\sigma}
\def\th{\theta}
\def\u{\mu}
\def\v{\nu}


\def\aEM{\alpha_{\scriptscriptstyle EM}}

\def\bar#1{\overline{#1}}
\def\Bv{{\bf B}_v}
\def\Bbarv{\bar{\bf B}_v}
\def\ccdot{\hbox{\kern-.1em$\cdot$\kern-.1em}}
\def\CDslash{\CD\hskip-0.65 em / \hskip+0.30 em}
\def\CPslash{{\scriptscriptstyle{C\hskip-0.2 em P}}\hskip-0.85 em
  {\scriptstyle \swarrow }}
\def\dash{{\> \over \>}} 		

\def\dN{{d_\N}}

\def\ecm{e \dash{\rm cm}}
\def\EM{{\scriptscriptstyle EM}}
\def\GeV{\>\, \rm GeV}
\def\gfive{\gamma^5}
\def\gtap{\raise.3ex\hbox{$>$\kern-.75em\lower1ex\hbox{$\sim$}}}

\def\ltap{\raise.3ex\hbox{$<$\kern-.75em\lower1ex\hbox{$\sim$}}}
\def\LX{{\Lambda_\chi}}
\def\Mb{M_\B}
\def\mbar{{\overline m}}
\def\MeV{\> {\rm MeV}}
\def\mpi{m_{\pi}}

\def\parenCM{{\scriptscriptstyle (\CM)}}

\def\parenM{{\scriptscriptstyle (M)}}
\def\parenth{{\scriptscriptstyle (\theta)}}
\def\parenzero{{\scriptscriptstyle (0)}}
\def\proj{{1 + \slash{v} \over 2}}

\def\qbar{\overline{q}}
\def\QCD{{\scriptscriptstyle QCD}}
\def\slash#1{#1\hskip-0.5em /}


\def\half{{1 \over 2}}
\def\quarter{{1 \over 4}}
\def\sixth{{ 1\over 6}}
\def\third{{1 \over 3}}

\def\twothirds{{2 \over 3}}


\newdimen\pmboffset
\pmboffset 0.022em
\def\oldpmb#1{\setbox0=\hbox{#1}%
 \copy0\kern-\wd0
 \kern\pmboffset\raise 1.732\pmboffset\copy0\kern-\wd0
 \kern\pmboffset\box0}
\def\pmb#1{\mathchoice{\oldpmb{$\displaystyle#1$}}{\oldpmb{$\textstyle#1$}}
      {\oldpmb{$\scriptstyle#1$}}{\oldpmb{$\scriptscriptstyle#1$}}}

\def\pib{{\pmb{\pi}}}


%

\nref\Wise{M. Wise, in Proceedings of the Banff Summer Institute, ed. A.N.
   Kamal and F.C. Khanna (World Scientific, 1988) p. 124.}
\nref\Baluni{V. Baluni, Phys. Rev. {\bf D19} (1979) 2227.}
\nref\Crewther{R.J. Crewther, P. Di Vecchia, G. Veneziano and E. Witten,
  Phys. Lett. {\bf B88} (1979) 123; (E) Phys. Lett. {\bf B91} (1980) 487.}
\nref\GasserLeutwylerI{J. Gasser and H. Leutwyler, Ann. Phys. {\bf 158}
  (1984) 142.}
\nref\GasserLeutwylerII{J. Gasser and H. Leutwyler, Nucl. Phys. {\bf
  B250} (1985) 465.}
\nref\Weinberg{S. Weinberg, in A Festschrift for I.I. Rabi (New
  York Academy of Sciences, New York, 1977) 185.}
\nref\CCWZ{S. Coleman, J. Wess and B. Zumino, Phys. Rev. {\bf 177} (1969)
  2239\semi
  C. Callan, S. Coleman, J. Wess and B. Zumino, Phys. Rev. {\bf 177} (1969)
  2247.}
\nref\JenkinsManoharI{
  E. Jenkins and A. Manohar, Phys. Lett. {\bf B255} (1991) 558.}
\nref\JenkinsManoharII{
  E. Jenkins and A. Manohar, Phys. Lett. {\bf B259} (1991) 353.}
\nref\Jenkins{E. Jenkins, Nucl. Phys. {\bf B368} (1992) 190.}
\nref\JenkinsManoharIII{
  E. Jenkins and A. Manohar, UCSD/PTH 91-30 (1991).}
\nref\Georgi{H. Georgi, Phys. Lett. {\bf B240} (1990) 447.}
\nref\HQETreviews{For reviews of HQET, see
  M. Wise, ``New Symmetries of the Strong Interactions'', Lectures presented
  at the Lake Louise Winter Institute, Feb 17-23 1991, CALT-68-1721;
  H. Georgi, ``Heavy Quark Effective Theory'', in Proc. of the Theoretical
  Advanced Study Institute (TASI) 1991, ed. R.K. Ellis, C.T. Hill and J.D.
  Lykken (World Scientific, Singapore, 1992) p. 589;
  B. Grinstein, ``Lectures on Heavy Quark Effective Theory'', in High Energy
  Phenomenology, Proceedings of the Workshop, Mexico City 1-12 July 1991,
  eds. R. Heurta and M.A. Perez, (World Scientific, Singapore),
  SSCL-Preprint-17.}
\nref\Shifman{M.A. Shifman, A.I. Vainshtein and V.I. Zakharov, Nucl.
  Phys. {\bf B166} 493.}
\nref\PDB{Review of Particle Properties, Phys. Rev. {\bf D45}, Part 2 (1992).}
\nref\Nuyts{J. Nuyts, Phys. Rev. Lett. {\bf 26} (1971) 1604.}
\nref\WessZumino{J. Wess and B. Zumino, Phys. Lett. {\bf B37} (1971) 95.}
\nref\Witten{E. Witten, Nucl. Phys. {\bf B223} (1983) 422.}
\nref\Bernreuther{W. Bernreuther and M. Suzuki, Rev. Mod. Phys. {\bf 63}
  (1991) 313.}
\nref\ChengI{H.-Y. Cheng, Phys. Rep. {\bf 158} (1988) 1, and references
  therein.}
\nref\Pich{A. Pich and E. de Rafael, Nucl. Phys. {\bf B367} (1991) 313.}
\nref\ChengII{H.-Y. Cheng, Phys. Rev. {\bf D44} (1991) 166.}

\nfig\tadpoles{One-loop $\pi^0$ and $\eta$ tadpole diagrams.  Large dots
denote CP violating vertices.  The cross in the $\pi^0 \pi^0 \eta$ graph
represents the $O(m_u-m_d)$ off-diagonal mixing term in the neutral
Goldstone boson propagator.}
\nfig\Pizerographs{CP violating graphs that contribute to $\pi^0 \to
\gamma \gamma$.  A virtual charged kaon runs around the loop.}
\nfig\NEDMgraphs{One-loop diagrams that contribute to the neutron
electric dipole moment.  Identical graphs with internal $\pi^-$ and $P_v$
propagators replaced by $K^+$ and $\Sigma^-_v$ lines are not pictured.}


\def\CITTitle#1#2{\nopagenumbers\abstractfont
\hsize=\hstitle\rightline{#1}
\vskip 1in\centerline{\titlefont #2~\footnote{$\hskip-0.2em^*\hskip-0.5em$}
  {\vskip-0.26in{\sevenrm Work supported in part by the U.S. Dept. of
  Energy under Contract no. DEAC-03-81ER40050.}}}
 \abstractfont\vskip .5in\pageno=0}

\CITTitle{{\baselineskip=12pt plus 1pt minus 1pt
  \vbox{\hbox{CALT-68-1842}\hbox{DOE RESEARCH AND}\hbox{DEVELOPMENT REPORT}}}}
  {Chiral Estimates of Strong CP Violation Revisited}
\centerline{Peter Cho}
\bigskip\centerline{\it California Institute of Technology, Pasadena, CA
  91125}

\vskip .6in
\centerline{\bf Abstract}
\bigskip

	The effects of the CP violating $\theta$ term in the QCD Lagrangian
upon low energy hadronic phenomenology are reconsidered.  Strong CP violating
interactions among Goldstone bosons and octet baryons are incorporated into
an effective chiral Lagrangian framework.  The $\theta$ term's impact upon the
decays $\eta\to\pi\pi$ and $\pi^0\to\gamma\gamma$ is then investigated but
found to be extremely small.  A refined model independent estimate of
nonanalytic contributions to the neutron electric dipole moment is also
determined using velocity dependent Baryon Chiral Perturbation Theory.  We
obtain the approximate upper bound $|\theta| < 4.5 \times 10^{-10}$.

\Date{12/92}

	CP invariance is known to be a very good but inexact symmetry of
the standard model.  Violations of this discrete symmetry have been
observed in neutral kaon decays and are hoped to be seen in bottom meson
phenomena such as $B-\bar{B}$ mixing \Wise.  CP violation in these weak
processes may be attributed to a complex phase in the Kobayashi-Maskawa
matrix.  However, this phase is not the only source of CP
violation within the minimal six quark standard model.  In the strong
interaction sector, instanton effects generate a $G \tilde{G}$ term which
preserves charge conjugation but breaks parity.  This topological term
enters into the QCD Lagrangian
\eqn\LQCD{\CL_{\QCD} = -\quarter G^{\u\v}_a G_{\u\v a} + \qbar i \CDslash q
  - \qbar M q + \th {g^2 \over 32 \pi^2} G^{\u\v}_a \tilde{G}_{\u\v a}}
with an undetermined coefficient $\th$ that represents a fundamental parameter
of the standard model.  In this letter, we investigate the effects of the
$\th$ term upon low energy hadronic phenomenology.

        To begin, we restrict the QCD Lagrangian to three light quarks
and take the mass matrix $M$ to be real and diagonal without loss of
generality:
\eqn\currmasses{M = \pmatrix{m_u && \cr & m_d & \cr && m_s \cr}.}
It is convenient to rotate the $\th$ parameter away from the $G \tilde{G}$
term in \LQCD\ and into the quark mass matrix which subsequently becomes
complex \refs{\Baluni,\Crewther}:
\eqn\LQCDnew{\eqalign{
  \CL_{\QCD} &\to -\quarter G^{\u\v}_a G_{\u\v a} + \qbar i \CDslash q
  -\qbar M q -i \th \mbar \, \qbar \gfive q \cr
 &= -\quarter G^{\u\v}_a G_{\u\v a} + \qbar_\L i \CDslash q_\L+
  \qbar_\R i\CDslash q_\R
    -\qbar_\L (M+i\th\mbar) q_\R - \qbar_\R (M-i \th\mbar) q_\L.\cr}}\
If any of the current masses in \currmasses\ equals zero, the reduced
quark mass
$$ \mbar = {m_u m_d m_s \over m_u m_d + m_d m_s + m_s m_u} $$
vanishes and the mass matrix remains real.  However in the real world, the
current quark masses have the small but nonvanishing values
\refs{\GasserLeutwylerI,\GasserLeutwylerII,\Weinberg}
\eqn\quarkmasses{(m_u,m_d,m_s) \simeq (5, 9, 181) \MeV}
which imply $\mbar \simeq 3.2 \MeV$.  So the $\th$ parameter disappears
from the topological term's coefficient and reemerges in
${\rm arg \bigl( det} (M+i\th\mbar) \bigr)=\th+O(\th^3)$.

        At low energies, it is useful to replace Lagrangian \LQCDnew\
which describes QCD in terms of fundamental quarks and gluons with
an effective Lagrangian of mesons and baryons.  In particular, the self
interactions of Goldstone bosons associated with the chiral symmetry
breakdown $G = SU(3)_\L \times SU(3)_\R \to H=SU(3)_{\L+\R}$ may be
analyzed in a nonlinear chiral Lagrangian framework \CCWZ.  The Goldstone
bosons appearing in the pion octet
\eqn\pionoctet{\pib = {1 \over \sqrt{2}}
  \pmatrix{ \sqrt{\half} \pi^0 + \sqrt{\sixth} \eta & \pi^+ & K^+ \cr
  \pi^- & - \sqrt{\half} \pi^0+\sqrt{\sixth}\eta & K^0 \cr
  K^- & \bar{K}^0 & - \sqrt{\twothirds}\eta \cr}}
enter into the chiral theory through the combinations
$\Sigma=e^{2i \pib/f}$ and $\xi=e^{i \pib/f}$ where $f=93 \MeV$
represents the pion decay constant to lowest order.  These
exponentiated fields transform under $G$ as
\eqna\Sigmatrans
$$ \eqalignno{\Sigma &\to L \Sigma R^\dagger & \Sigmatrans a \cr
\xi &\to L \xi U^\dagger(x) = U(x) \xi R^\dagger & \Sigmatrans b\cr} $$
where $L$ and $R$ denote global elements of $SU(3)_\L$ and $SU(3)_\R$
while local matrix $U(x)$ is implicitly defined by \Sigmatrans{b}.
Goldstone terms in the effective Lagrangian are constructed from $\Sigma$
and $\xi$ in a derivative expansion with respect to the chiral symmetry
breaking scale $\LX \approx 1 \GeV$.

        Incorporating baryons into this scheme seems problematic.
The momentum expansion fails for baryons since their masses are not small
compared to $\LX$.  However, a consistent approach for performing Baryon
Chiral Perturbation Theory has recently been developed by Jenkins and
Manohar \refs{\JenkinsManoharI{--}\JenkinsManoharIII} using ideas and
techniques familiar from the Heavy Quark Effective Theory
\refs{\Georgi,\HQETreviews}.  At low energies, a baryon may be regarded as
an almost on-shell heavy particle that travels along a straight worldline.
In this kinematic regime, a baryon has four-momentum $p=\Mb v+k$ where the
residual momentum $k$ is small compared to its rest mass $\Mb$.  As its
four-velocity $v$ is essentially unaffected by soft Goldstone boson absorption
or emission, the baryon can be described by the velocity dependent field
$$ B_v(x) = e^{i M_B \slash{v} v\cdot x} B(x) $$
which has the rest energy removed from its definition.  The use of such
fields allows one to formulate a well-behaved derivative expansion in
terms of the small quantity $k/\LX$. This is the central idea behind
velocity dependent Baryon Chiral Perturbation Theory.

        The leading contributions to the effective Lagrangian
$$ \CL_{eff} = \CL_\pi + \sum_v \CL_{B}(v) $$
appear in $d=4-\e$ dimensions and at the renormalization scale $\Lambda$ as~
\foot{In their original formulation of Baryon Chiral Perturbation Theory,
Jenkins and Manohar introduced spin operators $S^\u_v$ that act on the
velocity dependent baryon fields and incorporated them into the chiral
Lagrangian.  For the applications that we will consider, these spin
operators offer no real advantage over conventional gamma matrices.  We
therefore express the derivative interactions of Goldstone bosons with
baryons in our Lagrangian in terms of the latter rather than former
objects.}~$\hskip-0.25em^,\hskip-0.7em$~
\foot{The baryon decuplet may also be readily included into the low
energy Lagrangian \refs{\JenkinsManoharII{--}\JenkinsManoharIII}.
However, these fields do not influence CP violating baryon octet phenomena
such as the neutron electric dipole moment at leading nontrivial order.  We
consequently neglect them here.}
\eqna\Lzero
$$ \eqalignno{
\CL^\parenzero_\pi &= {\Lambda^{-\e} f^2 \over 4} \Tr (\partial^\u
  \Sigma^\dagger \partial_\u \Sigma) & \Lzero a \cr
\CL^\parenzero_\B(v) &= i \Tr \Bbarv v \ccdot \CD \Bv + D \Tr \Bbarv
 \gamma^\u \gfive \{ {\bf A}_\u,\Bv \} + F \Tr \Bbarv \gamma^\u \gfive
 [{\bf A}_\u,\Bv]. & \Lzero b \cr} $$
Pions derivatively couple to the baryon octet through the Goldstone vector
field
$${\bf V}^\u = \half (\xi^\dagger \partial^\u \xi
  + \xi \partial^\u \xi^\dagger)
  = {\Lambda^\e \over 2f^2} [ \pib, \partial^\u \pib]
  -{\Lambda^{2\e} \over 24 f^4} \Bigl[ \pib,
  \bigl[\pib, [\pib,\partial^\u \pib] \bigr] \Bigr] + O(\pib^6) $$
that resides within the covariant derivative
$$ \CD^\u \Bv = \partial^\u \Bv + [{\bf V}^\u, \Bv].$$
In addition, they communicate via the axial current
$$ {\bf A}^\u = {i \over 2}(\xi^\dagger \partial^\u \xi - \xi \partial^\u
  \xi^\dagger) = -{\Lambda^{\e/2} \over f} \partial^\u \pib +
  {\Lambda^{3\e/2} \over 6 f^3} \bigl[\pib,[\pib,\partial^\u \pib] \bigr]
  + O(\pib^5) $$
which explicitly appears in \Lzero{b}.  We adopt the best fit values
$D=0.61$ and $F=0.40$ reported in refs.~\refs{\Jenkins,\JenkinsManoharIII}
for the symmetric and antisymmetric axial current couplings.  We also include
electromagnetic interactions into the effective Lagrangian by gauging a
$U(1)_\EM$ subgroup of the global $SU(3)_\L \times SU(3)_\R$ symmetry group
and treating photons as external fields.  The partial derivatives acting
on hadron fields in \Lzero{}\ are then promoted to covariant derivatives
with respect to electromagnetism
$$ \eqalign{\partial^\u \Sigma &\to \partial^\u \Sigma - i \Lambda^{\e/2}
  e \CA^\u [ {\bf Q},\Sigma] \cr
\partial^\u \xi &\to \partial^\u \xi - i \Lambda^{\e/2}
  e \CA^\u [ {\bf Q},\xi] \cr
\partial^\u \Bv &\to \partial^\u \Bv - i \Lambda^{\e/2}
  e \CA^\u [ {\bf Q},\Bv] \cr} $$
where
$$ {\bf Q} = \pmatrix{\twothirds && \cr & -\third & \cr && - \third\cr}$$
denotes the quark charge matrix.

        The symmetry breaking effects of the quark mass matrix
$\CM\equiv M+i\th\mbar$ in the underlying QCD Lagrangian can be
systematically incorporated into the low energy theory via the spurion
procedure.  We first regard $\CM$ as a fictitious field and assign it the
transformation rules
$$ \eqalign{\CM &{\buildrel G \over \longrightarrow} L \CM R^\dagger \cr
\CM &{\buildrel {\rm CP} \over \longrightarrow} \CM^* \cr}$$
which preserve chiral symmetry and discrete CP in eqn.~\LQCDnew.  We
then construct $G$ and CP invariant terms in the chiral Lagrangian from
the hadron fields and matrix $\CM$.  Finally, we set the spurion to
its true constant value and thereby generate interactions that violate
chiral symmetry, flavor and CP in the effective theory.

        Working to linear order in $\CM$, we add the following terms
into the low energy Lagrangian:~
\foot{A common baryon mass shift has been extracted from the $\sigma$
term in $(8b)$ and implicitly absorbed into the average octet mass
parameter $\Mb$.}
\eqna\CMterms
$$ \eqalignno{
\CL^\parenCM_\pi &= {\Lambda^{-\e} f^2 \over 2} \mu \Tr
  (\Sigma\CM^\dagger+\CM\Sigma^\dagger) & \CMterms a \cr
\CL^\parenCM_\B (v) &= b_\D \Tr \Bbarv \{ \xi^\dagger \CM
  \xi^\dagger + \xi \CM^\dagger \xi, \Bv \} + b_\F \Tr \Bbarv
  [ \xi^\dagger \CM \xi^\dagger + \xi \CM^\dagger \xi, \Bv] \cr
  &\qquad + \o \Tr \Bigl( (\Sigma-1) \CM^\dagger+\CM(\Sigma^\dagger-1)\Bigr)
  \Tr \Bbarv \Bv.& \CMterms b\cr} $$
Decomposing the mass matrix into its real and imaginary parts, we obtain
CP conserving terms proportional to $M$
\eqna\Mterms
$$ \eqalignno{
\CL^\parenM_\pi &= {\Lambda^{-\e} f^2 \over 2} \mu \Tr
  M(\Sigma+\Sigma^\dagger) & \Mterms a \cr
\CL^\parenM_\B (v) &= b_\D \Tr \Bbarv \{ \xi^\dagger M
  \xi^\dagger + \xi M \xi, \Bv \} + b_\F \Tr \Bbarv [ \xi^\dagger M
  \xi^\dagger + \xi M \xi, \Bv] \cr
  &\qquad + \o \Tr M (\Sigma+\Sigma^\dagger-2) \Tr \Bbarv \Bv  &
  \Mterms b\cr} $$
and CP violating interactions linear in $\th\mbar$
\eqna\thetaterms
$$ \eqalignno{
\CL^\parenth_\pi &= -i{\Lambda^{-\e} f^2 \over 2} \mu\th\mbar \Tr
  (\Sigma-\Sigma^\dagger) \quad & \thetaterms a\cr
\CL^\parenth_\B (v) &= -i\th \mbar \Bigl[
  b_\D \Tr \Bbarv \{\Sigma-\Sigma^\dagger,\Bv\}
 +b_\F \Tr \Bbarv  [\Sigma-\Sigma^\dagger,\Bv]
 +\o \Tr(\Sigma-\Sigma^\dagger) \Tr \Bbarv \Bv \Bigr]. \cr
 && \thetaterms b\cr} $$
The leading terms in \Mterms{a,b}\ produce Goldstone boson masses and
split the baryon octet multiplet.  The parameters $\mu$,$b_\D$ and $b_\F$
as well as the mean baryon mass $\Mb$ can therefore be fixed by a fit to
the hadron mass spectrum using the assumed current quark
masses in \quarkmasses.  We find the approximate tree level values
\eqn\paramvalues{\eqalign{ \mu &= 1302 \MeV\cr b_\D &=0.18 \cr} \qquad
\eqalign{\Mb &= 1197 \MeV \cr b_\F &= -0.54. \cr}}

        Having set up the necessary formalism, we can now investigate
specific examples of strong CP violating phenomena.  We first consider
Goldstone boson processes.  Expanding pion Lagrangian \thetaterms{a},
we isolate the self interaction terms
\eqn\trilinear{
  \eqalign{\CL^\parenth_\pi &= - {4 \over 3} {\Lambda^{\e/2} \over f}
  \mu\th\mbar \Tr(\pib^3) + O(\pib^5) \cr
&= {\Lambda^{\e/2} \over f} \mu\th\mbar \Bigl[
  \pi^0 K^+ K^- - \pi^0 K^0 \bar{K}^0 + \sqrt{2} \pi^+ K^- K^0 + \sqrt{2}
  \pi^- K^+ \bar{K}^0 \cr
& \quad\qquad\qquad {2 \sqrt{3}\over 3} \eta \pi^+ \pi^-
  + {\sqrt{3}\over 3} \eta \pi^0 \pi^0 - {\sqrt{3} \over 3} \eta K^+ K^-
- {\sqrt{3} \over 3} \eta K^0 \bar{K}^0 - {\sqrt{3} \over 9} \eta^3 \Bigr]
 +O(\pib^5). \cr}}
{}From these trilinear vertices, one can easily compute the tree level rate
for the CP violating decay $\eta\to\pi\pi$ \refs{\Crewther,\Shifman}:
\eqn\etadecayrate{
  \Gamma(\eta \to \pi^+ \pi^-) = {1 \over 12 \pi} \Bigl(
  {\mu\th\mbar \over f} \Bigr)^2 {\sqrt{m_\eta^2 - 4 m_{\pi^+}^2} \over
  m_\eta^2} \approx 8.4 \times 10^{-2} \theta^2 \MeV.}
Comparison with the current experimental limit
$\Gamma(\eta \to \pi^+ \pi^-) _{exp} < 1.79 \times 10^{-6} \MeV$ \PDB\
implies that $|\th| < 4.6 \times 10^{-3}$.  However, much more stringent
bounds on $\th$ are available from electric dipole measurements.
So it is of greater use to insert $|\th| < 4.5 \times 10^{-10}$, the
constraint which comes from the neutron electric dipole analysis that we
will shortly present, into eqn.~\etadecayrate.  We then obtain
the prediction
$ \Gamma(\eta \to \pi^+ \pi^-)_{th} < 1.7 \times 10^{-20} \MeV.$
This theoretical bound is unfortunately 14 orders of magnitude smaller than
the experimental upper limit.  One may thus safely assume that the
$\th$ parameter will never be extracted from this Goldstone decay mode.

	The CP violating interactions in \trilinear\ also contribute to
loop processes.  The simplest one-loop diagrams involving the
Goldstone vertices are the $\pi^0$ and $\eta$ tadpoles displayed in
\tadpoles.  The graphs' $O(\CM)$ quadratic divergences automatically sum to
zero as a result of the vanishing tadpole renormalization condition
$$ \langle 0 | i \th \mbar \,\bar{q} \gfive q | \pi^a \rangle = 0 $$
which we implicitly invoked when rotating $\th$ into the quark mass
matrix \refs{\Crewther,\Nuyts}.  Terms that persist at higher orders in
$\CM$ along with tadpole interactions in the effective Lagrangian starting
at $O(\CM^2)$ may always be eliminated via an appropriate modification of
the $\th$ rotation.

	While neutral Goldstone bosons do not disappear directly into the
vacuum, they can decay into two photons.  As illustrated in \Pizerographs,
the process $\pi^0\to\g\g$ may proceed through kaon triangle and ``seagull''
graphs that violate CP.  A straightforward calculation yields the finite
and gauge invariant on-shell amplitude
\eqn\Pizeroamp{
 \eqalign{\CA \bigl( \pi^0(p+q) &\to\g (p) \g (q) \bigr) \CPslash = \cr
 & {2\mu\th\mbar e^2 \over 16\pi^2f}
 \Bigl[ {4 m_K^2 \over \mpi^2} \Bigl( \tan^{-1} {\mpi \over
 \sqrt{4 m_K^2 - \mpi^2}} \Bigr)^2 -1 \Bigr] \Bigl(g_{\u\v} -2 {p_\v q_\u
 \over \mpi^2} \Bigr) \varepsilon^\u(p) \varepsilon^\v(q) .\cr}}
{}From the form of this expression, we see that the graphs in \Pizerographs\
match onto an infinite string of nonrenormalizable operators that reduce
via the equations of motion to $\pi^0 F^{\u\v} F_{\u\v}$.  Since this operator
does not respect CP, it cannot interfere with the one pion-two photon vertex
in the Wess-Zumino action \refs{\WessZumino,\Witten}
\eqn\WZaction{
  W(\Sigma,\CA^\u) = \int d^4 x \Bigl\{-{N_c e^2 \over 48\pi^2 f}
  \pi^0 F^{\u\v} \tilde{F}_{\u\v}+\cdots \Bigr\}.}
The CP violating contribution to the decay rate
\eqn\Pizeroth{
 \Gamma(\pi^0 \to \g \g) \CPslash =
 {1 \over 64\pi} \Bigl( {\aEM\over\pi} \Bigr)^2
 \Bigl( {\mu\th\mbar\over f}\Bigr)^2 \mpi^{-1}
 \Bigl[ {4 m_K^2 \over \mpi^2} \Bigl( \tan^{-1} {\mpi \over
 \sqrt{4 m_K^2 - \mpi^2}} \Bigr)^2 -1 \Bigr]^2}
is therefore suppressed by two powers of $\th$ compared to its lowest order CP
conserving counterpart
\eqn\PizeroWZ{
 \Gamma(\pi^0\to\g\g)_{\W\Z} = {1 \over 64\pi} \Bigl( {\aEM\over\pi}\Bigr)^2
 \Bigl( {\mpi^2\over f}\Bigr)^2 \mpi^{-1}.}
Their ratio is consequently miniscule:
$$  {\Gamma(\pi^0\to\g\g)\CPslash\over\Gamma(\pi^0\to\g\g)_{\W\Z}} =
2.1 \times 10^{-6} \th^2 < 4 \times 10^{-25} ! $$

	The preceding Goldstone boson examples clearly demonstrate that one
cannot hope to observe strong CP violating effects unless they occur at
linear order in $\th$.  Fortunately, there is a well-known phenomenon in
the baryon sector which meets this criterion: the neutron electric dipole
moment (NEDM). Previous attempts to determine the magnitude of this
important observable have typically relied upon specific model calculations.
We will reconsider this problem in the context of velocity dependent
Baryon Chiral Perturbation Theory.  While our investigation is clearly
similar in spirit to the well-known current algebra NEDM analysis of Crewther,
Di Vecchia, Veneziano and Witten \Crewther, we believe that the effective
field theory approach is more transparent and systematic.
Furthermore, heavy hadron techniques are especially well suited for
studying the electric dipole moment $\dN$ since it is a static property.
Recall that $\dN$ is defined in terms of the form factor $F_3$ appearing in
the interaction Lagrangian
\eqn\dipole{\CL_\I = i {F_3(q^2)\over 2M_\N} \bar{N}(p') \o^{\u\v} \gfive
 N(p) F_{\u\v}}
evaluated at zero momentum transfer $(q=p'-p=0)$ \Bernreuther:
$$ \dN = {F_3(0)\over M_\N}.$$
The neutron's four-velocity is therefore conserved, and the assumptions
underlying the static hadron picture are genuinely satisfied.   So we view
the NEDM problem as a particularly nice application of Baryon Chiral
Perturbation Theory.

	We first enumerate the leading terms which may appear
in the low energy Lagrangian and directly contribute to $\dN$.  Group
theory counting indicates that there are ten independent terms which one can
form from the octet fields $\bar{B}_v$ and $B_v$, the photon
combination $eQ \o^{\u\v} \gfive F_{\u\v}$, and the parity-odd mass term
$\xi^\dagger\CM\xi^\dagger-\xi\CM^\dagger\xi$.  A convenient basis for
these dimension-six operators is given below:
\eqn\Ops{\eqalign{
O_1 &= {e \over\Lambda_\chi^2} \Tr \bigl( \Bbarv \o^{\u\v} \gfive F_{\u\v}
  \{{\bf Q},\Bv\} \bigr) \Tr(\CM\Sigma^\dagger-\Sigma\CM^\dagger) \cr
O_2 &= {e \over\Lambda_\chi^2} \Tr \bigl( \Bbarv \o^{\u\v} \gfive F_{\u\v}
  [{\bf Q},\Bv] \bigr) \Tr(\CM\Sigma^\dagger-\Sigma\CM^\dagger) \cr
O_3 &= {e \over\Lambda_\chi^2} \Tr \bigl( \Bbarv \o^{\u\v}\gfive F_{\u\v} \Bv
  \bigr) \Tr \bigl( {\bf Q} (\xi^\dagger\CM\xi^\dagger -
  \xi\CM^\dagger\xi)\bigr)\cr
O_4 &= {e \over\Lambda_\chi^2} \Tr \bigl( \Bbarv \o^{\u\v} \gfive F_{\u\v}
  (\xi^\dagger\CM\xi^\dagger - \xi\CM^\dagger\xi) \bigr)
  \Tr \bigl( {\bf Q} \Bv \bigr) \cr
O_5 &= {e \over\Lambda_\chi^2} \Tr \bigl( \Bbarv \o^{\u\v} \gfive F_{\u\v}
  {\bf Q} \bigr) \Tr\bigl( (\xi^\dagger\CM\xi^\dagger - \xi\CM^\dagger\xi) \Bv
  \bigr) \cr
O_6 &= {e \over\Lambda_\chi^2} \Tr \bigl( \Bbarv \o^{\u\v} \gfive F_{\u\v}
  [(\xi^\dagger\CM\xi^\dagger - \xi\CM^\dagger\xi) {\bf Q},\Bv] \bigr) \cr
O_7 &= {e \over\Lambda_\chi^2} \Tr \bigl( \Bbarv \o^{\u\v} \gfive F_{\u\v}
  [{\bf Q},\Bv] (\xi^\dagger\CM\xi^\dagger - \xi\CM^\dagger\xi) \bigr) \cr
O_8 &= {e \over\Lambda_\chi^2} \Tr \bigl( \Bbarv \o^{\u\v} \gfive F_{\u\v}
  [{\bf Q},(\xi^\dagger\CM\xi^\dagger - \xi\CM^\dagger\xi)\Bv] \bigr) \cr
O_9 &= {e \over\Lambda_\chi^2} \Bigl(
  \Tr \bigl( \Bbarv \o^{\u\v} \gfive F_{\u\v}
  \Bv [\xi^\dagger\CM\xi^\dagger - \xi\CM^\dagger\xi,{\bf Q}] \bigr) \cr
  & \qquad + \Tr \bigl( \Bbarv \o^{\u\v} \gfive F_{\u\v}
  [\xi^\dagger\CM\xi^\dagger - \xi\CM^\dagger\xi,{\bf Q} \Bv]\bigr) \Bigr) \cr
O_{10} &= {e \over\Lambda_\chi^2} \Tr \Bigl( \Bbarv \o^{\u\v} \gfive F_{\u\v}
  \Bigl\{{\bf Q},\bigl(\xi^\dagger\CM\xi^\dagger - \xi\CM^\dagger\xi-\third
  \Tr(\xi^\dagger\CM\xi^\dagger - \xi\CM^\dagger\xi) \bigl)\Bv
  \Bigr\}\Bigr) .\cr}}
Of these ten operators, only $O_1$ affects the electric dipole moment
of neutral baryons.  Its impact upon $\dN$ is unknown however since $O_1$
appears in the effective Lagrangian with an {\it a priori} undetermined
coefficient.  In principle, the coefficients of all the operators in \Ops\
could be fixed from Goldstone boson-photoproduction data.  But in practice,
the challenge of performing such a fit appears formidable.

	As first noted in ref.~\Crewther, the most important
contributions to the neutron's dipole moment actually do not come from tree
level composite operators but rather from one-loop Goldstone boson
diagrams.  Such graphs generate infrared log renormalizations of $O_1$ which
diverge in the chiral limit.  Their nonanalytic dependence on $\CM$ is exactly
calculable unlike the analytic dependence of $O_1$'s coefficient.
The nonanalytic terms consequently provide a rough but useful
order-of-magnitude estimate for $\dN$.

        There are a number of CP violating diagrams that enter into the
neutron-neutron-photon 1PI Green's function $\Gamma^{(NN\g)}$ at one-loop
order.  But only those listed in \NEDMgraphs\ yield chiral log corrections
to the NEDM.  In addition, these graphs contain fractional power terms
starting at $O(\CM^{3/2})$ which result from the mass splitting between the
external neutron and intermediate baryon.  Summing the diagrams with
internal $\pi^-$ and $P$ propagators as well as $K^+$ and $\Sigma^-$ lines,
we obtain
\eqn\Graphsum{
  \eqalign{\Gamma^{(NN\g)} = {4 \th\mbar e \over 16 \pi^2 f^2} \Biggl\{
  & (D+F)(b_D+b_F) \Bigl[ \log {\Lambda^2 \over \mpi^2}- {\pi (M_\P-M_\N) \over
  \sqrt{\mpi^2 - (M_\P-M_\N)^2}} \Bigr] \cr
-&(D-F)(b_D-b_F) \Bigl[\log{\Lambda^2  \over m_K^2} -{\pi (M_\Sigma-M_N) \over
  \sqrt{m_K^2 - (M_\Sigma-M_N)^2}} \Bigr] \Biggr\} \cr
\times \bar{N}_v(k') & (v^\u \g^\v - v^\v \g^\u) \gfive (k'-k)_\v N_v(k)
  + \cdots .\cr}}
The Dirac algebra identity
$$ \proj (v^\u \g^\v - v^\v \g^\u) \gfive\proj = -i \proj \o^{\u\v} \gfive
  \proj $$
transforms the velocity dependent expression into a manifest dipole
operator.  Numerically evaluating eqn.~\Graphsum\ at the scale $\Lambda=
\LX = 1 \GeV$, we find for its coefficient
$$ \Gamma^{(NN\g)} = \Bigl[ (2.76-0.08) \times 10^{-16} \th \ecm
\Bigr] i \bar{N}_v(k') \o^{\u\v} \gfive (k'-k)_\v N_v(k) + \cdots $$
where the pion and kaon contributions are separately displayed. The
large disparity between these terms stems in part from their $SU(3)$
couplings and infrared logarithms.  But more importantly, the discrepancy
results from the near cancellation between the $\log\CM$
and $\CM^{3/2}$ terms in the strange virtual hadron graphs.
The total nonanalytic contribution to the neutron electric dipole
moment is thus given by
$$ \dN = (2.68 \times 10^{-16} \th) \ecm. $$
Comparing with the current experimental upper limit
$|\dN| < 1.2 \times 10^{-25}\ecm$ at $95 \%$ CL \PDB, we deduce
$|\th| < 4.5 \times 10^{-10}.$

	In conclusion, the results from our chiral Lagrangian investigation
of the $\th$ term's effect upon low energy hadron phenomenology are in
basic accord with earlier findings \ChengI.  Its virtue is therefore not
novelty but rather simplicity.  Of course, as in any effective field theory
analysis, the results are model independent and can be systematically
improved by retaining higher order terms in the derivative expansion.  In
particular, nonanalytic $O(1/M_\B)$ corrections to the NEDM could be
determined.  However given the uncertainty in the analytic contributions to
$|\dN|$, such further refinement of our estimate for $\th$ is not of much
practical importance.

\bigskip
\centerline{\bf Acknowledgements}
\bigskip

	It is a pleasure to thank Sandip Trivedi and Mark Wise for many
helpful discussions.

\bigskip\bigskip\bigskip\bigskip\noindent
{\it Added note}: After completion of this work, we learned that results
similar to those presented here have been reported in ref.~\Pich.  See
also ref.~\ChengII.

\listrefs
\listfigs
\bye